\documentclass[letter,twocolumn,letterpaper]{jpsj2}

\title{Charge ordering in $\theta$-(BEDT-TTF)$_2$X materials}

\author{R.T. Clay$^{1,2}$, S. Mazumdar$^1$, and D.K. Campbell$^3$}
\inst{ $^1$ Department of Physics, University of Arizona Tucson, AZ
85721\\ 
$^2$ Cooperative Excitation Project ERATO, Japan Science and
Technology Corporation (JST), Tucson, AZ 85721 \\ 
$^3$ Departments of
Electrical and Computer Engineering and Physics, Boston University,
Boston, MA 02215} 
\recdate{\today} 
\abst{We investigate theoretically charge ordered states on the
anisotropic triangular lattice characteristic of the
$\theta$-(BEDT-TTF)$_2$X materials.  Using exact diagonalization
studies, we establish that the charge order (CO) pattern corresponds
to a ``horizontal'' stripe structure, with ...1100... CO along the two
directions with larger electron hopping (p-directions), and
...1010... CO along the third direction (c-direction). The CO is
accompanied by co-operative bond dimerizations along all three
directions in the highest spin state.  In the lowest spin state bonds
along the p-directions are tetramerized.  Our theory explains the
occurence of a charge-induced high temperature transition as well as a
spin gap transition at lower temperature.}

\kword{charge ordering, charge-density wave, BEDT-TTF, spin gap, superconductivity}

\begin{document}
\maketitle

Ground states involving spatially inhomogeneous charge distributions
have in recent years been observed in a wide range of novel electronic
solids, particularly in reduced dimensions.  In the case of the
organic charge-transfer solids (CTS), it has been suggested that the
appearance of charge-order (CO) may be related to superconductivity
\cite{Mazumdar00a,Merino01a}.  In the present Letter, we focus on CO
within one particular class of the CTS, the $\theta$-(ET)$_2$X
materials (here ET is short for BEDT-TTF).  The $\theta$-(ET)$_2$X
materials are particularly interesting for several reasons: (i) many
detailed experimental studies have been performed; (ii) some members
(e.g., the $X=I_3$ salt \cite{Tamura90a,Kobayashi86a}) of this class
of CTS are superconducting; and (iii) the structure of the ET
conducting layer is fully two dimensional (2D), implying that standard
single-particle theories of broken symmetry based on nesting are
invalid. In this Letter we present exact numerical many-body
calculations establishing that the CO within the $\theta$-(ET)$_2$X
materials is a combined bond-charge density wave (BCDW) state
exhibiting a 2D extension of the ``...1100...'' type CO and BCDW that
we have previously shown exist in a wide variety of quasi-one
dimensional CTS \cite{Mazumdar00a}.  The 2D bond distortions of this
state also explain the spin gap seen at low temperatures in the
$\theta$-(ET)$_2$X.

\begin{figure}[tb]
\centerline{\resizebox{2.3in}{!}{\includegraphics{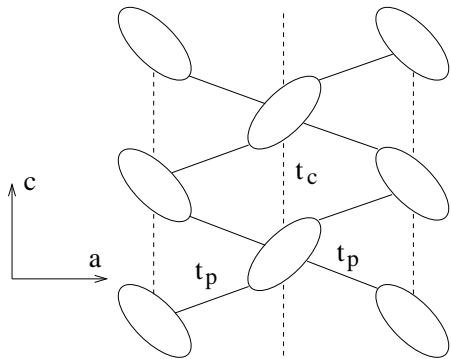}}}
\caption{Schematic structure of the $\theta$-(ET)$_2$X conducting layers,
consisting of the ET cationic molecules.
Solid lines correspond to stronger hopping $t_p$, dashed lines to weaker
hopping $t_c$.} 
\label{ETfig}
\end{figure}
\begin{figure}
\centerline{\resizebox{3.0in}{!}{\includegraphics{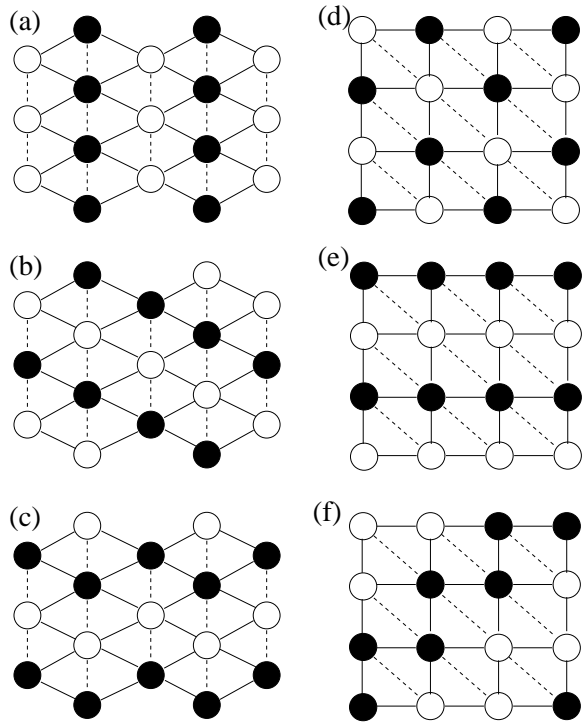}}}
\caption{Different possible CO patterns in the $\theta$-ET
materials. (a) vertical stripe (b) diagonal stripe, and (c) horizontal
stripe.  Filled (unfilled) circles correspond to molecular site with
greater (smaller) charge density.  Corresponding stripe orders within
a rotated rectangular lattice with hopping $t_p$ along the $x$ and
$y$-axes, and $t_c$ along the $-x+y$ diagonal are shown in (d), (e),
and (f). The hopping along the other diagonal is zero.}
\label{stripes}
\end{figure}

The structure of the 2D conducting layer in the $\theta$-type
materials is shown in Fig.~\ref{ETfig}.  We briefly review the
experimental results, focusing on the most widely investigated
$\theta$-(ET)$_2$MM$'$(SCN)$_4$ materials.  Here M is any of the
monovalent cations Rb, Tl and Cs, while M$'$ is one of the divalent
cations Co and Zn.  Slowly cooled $\theta$-RbZn exhibits a
metal-insulator (M-I) transition at 190 K that is accompanied by a
period doubling along the c-direction \cite{Mori98a,Mori98b}.  Direct
evidence for CO was found from NMR studies in this material, although
the actual pattern of CO was not discussed \cite{Miyagawa00a}.  From
analysis of the $^{13}$C-NMR line shapes, Chiba et al. have concluded
that the CO in $\theta$-RbZn corresponds to the {\it horizontal}
stripe structure (see Fig.~\ref{stripes}(c)) \cite{Chiba01a}. A
similar conclusion was reached also from the analysis of the
reflectance spectrum \cite{Tajima00a}. The 190 K transition has no
effect on the magnetic susceptibility, but a spin-singlet phase is
obtained in this material below 30 K
\cite{Mori98b,Nakamura99a,Miyagawa00a}.  The lower temperature
transition has been widely referred to as a spin-Peierls (SP)
transition \cite{Mori98b}.

In $\theta$-RbCo, the M-I transition also occurs at 190 K and again
there is a second transition (below about 40 K) to a spin singlet
state\cite{Mori98a,Mori98b}.  Crystal structure analysis gave clear
evidence of dimerization along the c-direction \cite{Mori98a} below
this temperature.  Importantly, since the dimerization in the
c-direction has already occurred at the M-I transition {\it this
dimerization cannot itself be the origin of the spin gap.}

The $\theta$-TlZn material is already semiconducting at room
temperature, but a resistive anomaly is seen at 165 K that may be
accompanied by a lattice dimerization along the c-direction
\cite{Mori98a}.  Evidence for short-range dimerization has been also
found in $\theta$-CsCo at the resistive transition, which, however,
occurs at a much lower temperature of 20 K
\cite{Watanabe99a,Nogami99a}. In $\theta$-CsZn, NMR measurements have
suggested a phase transition to a CO state near 20K
\cite{Nakamura00a}.  Finally, a different $\theta$-(ET) material,
$\theta$-(ET)$_2$Cu$_2$(CN)[N(CN)$_2$]$_2$, is also of interest in the
present context. This material shows a semiconductor to semiconductor
transition at 220 K that is accompanied by a weak lattice dimerization
along the c-direction \cite{Komatsu95a}.  No magnetic anomaly is seen
at this temperature. Below about 50 K there is a rapid drop in the
spin susceptibility, while the lattice dimerization along the
c-direction gets stronger \cite{Komatsu95a}.

Summarizing the experimental data, the most notable features 
are (a) the existence of two transitions, one 
involving charge at high temperatures and the other involving spin at
low temperatures; (b) where clear evidence exists, the horizontal
stripe charge order (see Fig.~\ref{stripes}(c)) is found; and (c)
lattice dimerization along the c-direction occurs already at the M-I
transition and therefore involves the charge degrees of freedom.  We
emphasize that the dimerization along the c-direction must necessarily
involve distortions of intermolecular distances, and not just charge
modulation, since ordinary (as opposed to anomalous) X-ray scattering
experiments determine only bond modulations and not site charge
modulations \cite{Ravy00a}.  The origin of the magnetic transition
remains an open question, since the dimerization that appears at high
temperature cannot itself produce a spin gap at lower temperature.

Both Seo \cite{Seo00a} and McKenzie et al. \cite{McKenzie01a} have
proposed models for the charge ordering in $\theta$-(ET).  Following
these authors and our own previous work \cite{Mazumdar00a}, we
investigate the electronic properties of the ET layer using the
extended Hubbard model: 
\begin{subequations}
\begin{align}
H &= H_0 +H_{ee} \\
H_0 &= -\sum_{\langle ij \rangle,\sigma}t_{ij}c_{i,\sigma}^\dagger 
c_{j,\sigma} \\
H_{ee} &= U\sum_{i}n_{i,\uparrow}n_{i,\downarrow} + 
V\sum_{\langle ij \rangle}n_{i}n_{j} 
\label{eqn-extHub}
\end{align}
\end{subequations}
In the above, $i, j$ are site indices, $\langle ... \rangle$
implies nearest neighbors, $n_j=n_{j,\uparrow}+n_{j,\downarrow}$, and
$\sigma$ is spin.  The average number of electrons (or holes) per site
$\rho$ equals $\frac{1}{2}$ for these $\frac{1}{4}$-filled materials.
In accordance with extended H\"uckel calculations \cite{Mori98a}, we
consider $t_c <$ 0, $t_p >$ 0, and $|t_c| << |t_p|$ within Eq.~1.
However, we expect $V_p$ and $V_c$ to be roughly the same size, since
while the hopping integrals depend upon the relative orientations of
the molecules, the Coulomb integrals depend only on the distances
between them. Hence one must include both $V_p$ and $V_c$.  We have
not included explicitly either intra- or inter-site electron-phonon
couplings in Eq.~1.  We will argue below that the signatures as well
as the effects of such couplings can be anticipated from accurate
solutions to the static extended Hubbard Hamiltonian for the ET
lattice.

Seo \cite{Seo00a} has examined three possible CO patterns within the
lattice of Fig.~\ref{ETfig} at quarter-filling.  All of these involve
``stripes'', which can be {\it vertical}, {\it diagonal}, or {\it
horizontal} stripes, as shown in Figs.~\ref{stripes} (a), (b), and
(c).  The lattice of Fig.~\ref{ETfig} may also be thought of as a
square lattice, with the essential proviso that hopping and Coulomb
interactions along one of the diagonals must be included.  In the
square lattice picture, the bonds along the two p-directions occur
along the $x$ and $y$ directions, respectively, and the intrastack
bond along the c-direction in this case corresponds to one of the two
diagonals of the square lattice.  In this picture, the vertical stripe
structure becomes the simple checkerboard pattern of site
occupancy. Similarly, the diagonal stripe structure has alternate
fully occupied and vacant rows along the $x$ direction.  Hence, both
the vertical and diagonal stripe charge orderings are based on
...1010... CO in different lattice directions.
In contrast, the horizontal stripe corresponds to {\it a 2D version of
...1100... CO} \cite{Mazumdar00a}, with the ...1100... orderings
occurring along both $x$ and $y$ directions {\it and} ..1010... CO
occurring along the diagonal correspond to the original c-axis. In
nearly all cases that Seo studied using Hartree-Fock (HF)/Hartree
solutions of Eq.~1, he found the vertical stripe to be the ground
state, competing occasionally with the diagonal stripe. Only by
introducing unequal intersite Coulomb correlations $V_p < V_c$ and
bond alternation along the c-direction did Seo find a CO pattern that
is not vertical.  In this specific case, the ground state was {\it
either} the diagonal stripe {\it or} the horizontal stripe, but these
were so close in energy that it was difficult to distinguish between
the two cases (see Fig.~15 of reference \cite{Seo00a}).  McKenzie et
al. studied the square lattice {\it without $t_c$ or $V_c$} in the
$U\rightarrow\infty$ limit within the slave-Boson approximation
\cite{McKenzie01a}, and for finite $U$ with exact diagonalization
\cite{Calandra02a}.  Ignoring $V_c$ obviously strongly favors the
checkerboard pattern CO in the square lattice (see
Fig.~\ref{stripes}(d)) which corresponds to the vertical stripe order
(see Fig.~\ref{stripes}(a)), and unsurprisingly these authors find
vertical stripe CO in their model and assert that this is what is
observed in the $\theta$-ET CTS.

Several important issues remain unresolved from these studies.
Seo's work was done within the mean field approximation \cite{Seo00a},
which is known to give erroneous behavior for the Hamiltonian of
Eq.~1; for instance, in 1D, the critical $V$ for ...1010... CDW
formation within HF theory is much too small, and the behavior of this
transition as a function of $U$ is incorrect
\cite{Shibata01a,Clay01a}. McKenzie et al.'s calculation
\cite{McKenzie01a,Calandra02a} ignores the Coulomb interaction $V_c$
completely.  While the hopping $t_c$ is indeed small and might
legitimately be neglected, $V$ is of similar magnitude in both $p$ and
$c$ directions (because the distances are similar), and $V_c/t_c$ is
actually much larger than $V_p/t_p$.  Importantly, previous
theories\cite{Seo00a,McKenzie01a,Calandra02a} have failed to explain
(a) the c-axis dimerization (since the vertical stripe CO exhibits
neither charge nor bond dimerization in the c direction), and (b) the
low temperature spin gap transition.  It is argued in reference
\cite{Seo00a} that inclusion of e-ph interactions would lead to bond
alternation among the occupied sites along individual 1D stripes
(i.e. along the ...1-1-1...  bonds in Fig.~\ref{stripes}), as in a
normal 1D spin-Peierls transition.  Alternatively, Seo and Fukuyama
have proposed a frustrated spin model to account for the spin gap
\cite{Seo98a}.  However, it is not clear how the frustrated and CO
states could coexist.

\begin{figure}[tb]
\centerline{\resizebox{2.2in}{!}{\includegraphics{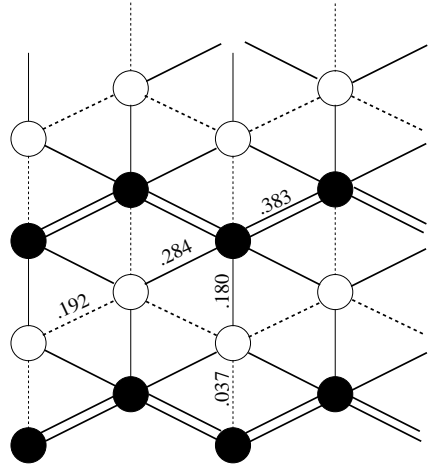}}}
\caption{The periodic 16 site lattice investigated for all three
stripe CO orders. Charge densities and bond orders within the
horizontal stripe order are shown for the ground spin singlet state
for the parameters $|t_p|=0.14$ eV, $|t_c|=0.01$ eV, $U=0.7$ eV, 
$V = 0.35$ eV, and $|\epsilon_i| = 0.01$ eV. Note the bond dimerization
along the c-direction and the bond tetramerization along the
p-directions. A similar calculation for the ferromagnetic state shows
the same bond dimerization along the c-direction, but instead of
tetramerization, bonds are now dimerized even along the p-directions
(see text).
Differences in the bond orders occur spontaneously as a result of
charge ordering even though hopping integrals are uniform along $t_c$
and $t_p$ directions in our calculations.}
\label{bond-order}
\end{figure}

To go beyond the previous approximations and assumptions, we have
performed exact diagonalization studies for a 16-site lattice with 8
electrons, for the lattice structure of Fig.~\ref{ETfig} (see
Fig.~\ref{bond-order} for the actual finite-size lattice that was
investigated).  We choose boundary conditions that are periodic along
$t_c$ and both the $t_p$ directions, such that all three CO patterns
of Fig.~\ref{stripes} are well-defined and can be studied in an
unbiased manner (i.e., the classical energies of the three stripe
patterns obtained by setting $t_{ij} = 0$, are identical).  We
consider all hoppings $t_p$ to be identical, and similarly all
hoppings $t_c$ are taken to be the same.  Since finite periodic
systems do not exhibit broken symmetry, we follow our previously
established procedure\cite{Mazumdar00a} and add a site energy
component $\sum_{i}\epsilon_in_i$ to the Hamiltonian of Eq.~(1), where
$\epsilon_i$ is negative for the ``occupied'' sites and positive for
the ``unoccupied'' sites.  This is equivalent to including on-site
electron-phonon (e-ph) coupling effects (within the classical
approximation), with {\it fixed} spring constants for all three stripe
structures.  We then calculate the lowest energy corresponding to each
specific CO for the smallest $\epsilon_i$ that still gives measurable
energy differences between the three stripe structures.  As the
spring-constant phonon energy is the same for on-site e-ph
interactions for all stripe orders considered, the electronic energies
in the static limit for small $\epsilon_i$ provide correct measures of
the relative stabilities of the different CO patterns in the limit of
$0^+$ on-site e-ph coupling.

We have chosen parameters similar to those in reference \cite{Seo00a},
and our results below are for $U$ = 0.7 eV, $t_c$ = --0.01 eV and
$t_p$ = 0.14 eV (note the overall minus sign in $H_0$ in Eq.~(1)),
with $V$ ranging from 0.15 -- 0.35 eV and $|\epsilon_i|$ = 0.01 eV.
Our range of $V$ covers the entire realistic region $V<\frac{1}{2}U$,
given the fixed value of $U$ = 0.7 eV.  We have confirmed that
increasing $t_c$ by a factor of 2 -- 3 or decreasing $|\epsilon_i|$
further does not lead to any modification of the phase diagram shown
below.
As we have discussed
extensively in our previous work \cite{Mazumdar00a}, finite-size
calculations tend to {\it overemphasize} the ...1010... CO and {\it
under-emphasize} the ...1100... CO.  Thus the relative energy of the
horizontal stripe CO in the present case is an {\it upper limit}, and
the actual phase space region where this CO dominates in the
thermodynamic limit is {\it larger} than what we find below.

\begin{figure}[tb]
\centerline{\resizebox{3in}{!}{\includegraphics{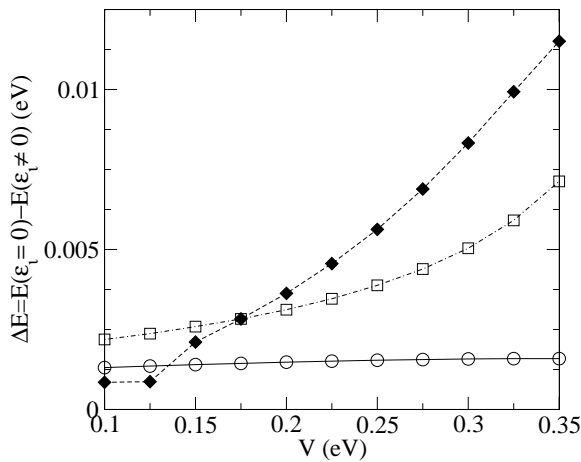}}}
\caption{The energy gained upon stripe formation, for the three stripe
patterns of Fig.~\ref{stripes}. Circles, squares and diamonds
correspond to vertical, diagonal and horizontal stripe patterns,
respectively. For $V >$ 0.18 eV, the ground state has the horizontal
stripe CO.}
\label{stripe-energy}
\end{figure}

In Fig.~\ref{stripe-energy} we have plotted the energy gained upon
stripe formation, $\Delta E = E(\epsilon_i=0) - E(\epsilon_i\neq 0)$
for the vertical, diagonal and horizontal stripe phases, respectively,
versus $V$ for the parameters given above.  We find that in all cases
the ground state contains either the diagonal stripe or the horizontal
stripe. Importantly, the horizontal stripe CO dominates over a broad
region of the phase space ($V>$0.18 eV), despite the isotropic
intersite Coulomb integrals and uniform $t_c$.

In addition to the energies, we have calculated the charge densities
$n_j$ and bond orders, the latter defined as the expectation value of
$b_{ij}\equiv\sum_\sigma (c^\dagger_{i\sigma}c_{j\sigma} +
c^\dagger_{j\sigma}c_{i\sigma})$.  We find that the order parameter
$\Delta n$ here can be very large ($\sim$ 0.3),
for $|\epsilon_i|=0.01$ eV, although this value will
depend on the actual strength of the on-site e-ph coupling.  However,
the bond orders in a given direction are {\it uniform} within both the
vertical and diagonal stripe phases. On the other hand, within the
horizontal CO, the bond orders alternate along the c-direction, while
along the two p-directions there occurs a tetramerized
periodicity. This is shown in Fig.~\ref{bond-order}.  There is thus a
strong tendency to {\it spontaneous bond distortions in the horizontal
CO phase in all three directions:} the differences in the bond orders
in Fig.~\ref{bond-order} suggest that for arbitrarily small
inter-site e-ph coupling there would occur bond dimerizations
along the c-direction and bond tetramerizations along the two diagonal
p-directions in the horizontal stripe phase.  {\it The tendency to
bond distortion in the $\theta$-ET lattice is a consequence of the
co-operative BCDW nature of the horizontal stripe phase}
\cite{Mazumdar00a}.  If finite intra- and/or inter-site e-ph couplings
are included, the cooperative nature of the BCDW accompanying
horizontal stripe phase will further lower its energy. It is then
likely that even in the small $V$ region of Fig.~\ref{stripe-energy}
that the horizontal stripe will dominate.

We have performed similar calculations for the {\it ferromagnetic}
spin configuration (total spin S = S$_{max}$ = 4), and have determined
that the bond dimerization in the c-direction persists even here,
although the magnitude of the dimerization is smaller.  Thus, lattice
dimerization in the c-direction is expected at the M-I transition, as
soon as high spin states of the horizontal stripe order are formed.
In contrast to the bond tetramerizations along the p-directions in the
low spin state, however, {\it the bonds in the p-directions in the
high spin state are dimerized}: the 1-1 and 0-0 bonds are the same,
and the 0-1 and 1-0 bonds are the same, but the two pairs are
different.  From our previous work \cite{Mazumdar00a}, there can be no
spin gap in this S = S$_{max}$ state with bond dimerizations, but spin
gaps will appear in the S=0 ``p-tetramerized'' state along both
p-directions. We further note that the horizontal stripes can {\it
not} be viewed as 1D half-filled systems, because the ``unoccupied''
sites in this phase actually have small spin components that are
coupled to the large spin components on the occupied sites, rendering
the spin behavior in this phase fundamentally 2D.

In reference \cite{Seo00a} Seo found that explicit inclusion of a bond
dimerization along the c-direction (in the form of alternating $t_c$)
lowers the relative energy of the horizontal stripe phase (though not
necessarily giving this as the ground state).  The present exact
calculations indicate that the {\it lattice dimerization in the
c-direction is a consequence of the horizontal stripe order, rather
than being the cause of it.}

Our conclusions can be summarized as follows.  We have shown that
exact many-body calculations incorporating all relevant
electron-electron interactions predict horizontal stripe CO in the
$\theta$-ET materials for a substantial region of parameter
space. This result contradicts the previous results of Seo
\cite{Seo00a} and McKenzie et al \cite{McKenzie01a,Calandra02a} 
and is consistent with experiments.
In both the ferromagnetic state and the low spin state the CO produces
bond dimerization along the c-direction, with the amplitude of the
dimerization larger in the low spin state. This explains the observed
appearance of c-axis dimerization at the M-I
transition\cite{Mori98a,Mori98b,Watanabe99a,Nogami99a,Nakamura00a,Komatsu95a}
as well as its increase at low temperatures\cite{Komatsu95a}, 
where the free energy is dominated by the low spin state.
In the two p-directions, bonds are dimerized in the
ferromagnetic state 
but tetramerized in the low spin state.
Our work thus explains naturally two distinct
transitions in the $\theta$-ET materials, a high temperature
transition \cite{Mori98a,Watanabe99a,Nogami99a,Nakamura00a,Komatsu95a}
that occurs as soon as the charge-occupancies correspond to the
horizontal stripe CO, and a low temperature magnetic transition
\cite{Mori98b,Miyagawa00a,Nakamura99a} 
The spin gap transition
is different from the usual SP transition, which is due to bond
alternation in the 1D half-filled spin system.  The formation of a
spin gap in 2D in the absence of frustration is a unique feature of
the strongly correlated 2D $\frac{1}{4}$-filled band.  Finally, our
results raise questions regarding the theory of charge-fluctuation
mediated superconductivity proposed by Merino et al. \cite{Merino01a},
as this work is based on the incorrect assumption of a checkerboard
vertical stripe CO.

S.M. and D.K.C. acknowledge partial support from the US NSF. Numerical
calculations were done in part at the NCSA.


\begin{thebibliography}{99}
\bibitem{Mazumdar00a} 
S. Mazumdar {\it et al.}, Phys. Rev. Lett. {\bf 82}, 1522 (1999);
S. Mazumdar,  R.~T. Clay, and D.~K. Campbell,
Phys. Rev. B {\bf 62}, 13400 (2000).
\bibitem{Merino01a} J. Merino and R.~H. McKenzie, Phys. Rev. Lett.
{\bf 87}, 237002 (2001).
\bibitem{Tamura90a} M. Tamura {\it et al.}, J. Phys. Soc. Jpn. {\bf 59},
1753 (1990).
\bibitem{Kobayashi86a} H. Kobayashi {\it et al.}, Chem. Lett. 789 (1986).
\bibitem{Mori98a} H. Mori {\it et al.} 
Bull. Chem. Soc. Jpn. {\bf 71}, 797 (1998).
\bibitem{Mori98b} H. Mori, S. Tanaka, and T. Mori, Phys. Rev. B.
{\bf 57}, 12023 (1998).
\bibitem{Miyagawa00a} K. Miyagawa, A. Kawamoto, and K. Kanoda, Phys.
Rev. B {\bf 62}, R7679 (2000).
\bibitem{Chiba01a} R. Chiba {\it et al.}, Synth. Metals {\bf 120},
919 (2001); R. Chiba {\it et al.}, J. Phys. Chem. Solids
{\bf 62}, 389 (2001).
\bibitem{Tajima00a} H. Tajima {\it et al.}, 
Phys. Rev. B {\bf 62}, 9378 (2000).
\bibitem{Nakamura99a} T. Nakamura {\it et al.}, Synth. Metals {\bf 103},
1898 (1999).
\bibitem{Watanabe99a} M. Watanabe {\it et al.}, 
J. Phys. Soc. Jpn. {\bf 68}, 2654 (1999).
\bibitem{Nogami99a} Y. Nogami {\it et al.}, Synth. Metals {\bf 102},
1778 (1999).
\bibitem{Nakamura00a} T. Nakamura {\it et al.},
J. Phys. Soc. Jpn. {\bf 69}, 594 (2000).
\bibitem{Komatsu95a} T. Komatsu {\it et al.}, Bull. Chem. Soc. Jpn.
{\bf 68}, 2233 (1995).
\bibitem{Ravy00a} S. Ravy and J.~E. Lorenzo, cond-mat/0005044.
\bibitem{Seo00a} H. Seo, J. Phys. Soc. Jpn. {\bf 69}, 805 (2000).
\bibitem{McKenzie01a} R.~H. McKenzie {\it et al.},
Phys. Rev. B {\bf 61}, 085109 (2001).
\bibitem{Calandra02a} M. Calandra, J. Merino, and R.~H. McKenzie,
cond-mat/0202219.
\bibitem{Shibata01a} Y. Shibata, S. Nishimoto, and Y. Ohta, Phys.
Rev. B {\bf 64}, 235107 (2001).
\bibitem{Clay01a} R.~T. Clay, S. Mazumdar, and D.~K. Campbell,
cond-mat/0112278.
\bibitem{Seo98a} H. Seo and H. Fukuyama, J. Phys. Soc. Jpn. {\bf 67},
1848 (1998).
\end{thebibliography}
\end{document}